# THE REPULSIVE FORCE IN CONTINUOUS SPACE MODELS OF PEDESTRIAN MOVEMENT

## Bernhard Steffen, Armin Seyfried

*Jülich Supercomputing Center (JSC), Forschungszentrum Jülich GmbH, 52425 Jülich, Germany*
*email: b.steffen@fz-juelich.de, a.seyfried@fz-juelich.de*

**Abstract**

Pedestrian movements can be modeled at different degrees of detail. While flux models (Predeshensky/Milinski 1971) and cellular automata models (Schreckenberg 2002) give answers to some important questions and are fast and easy to use, continuous space modeling has the potential of full flexibility in geometry and realistic description of individual movements in arbitrary fine resolution. While the acceleration forces in these models are known with good reliability, there is no agreement on the repulsive forces, not even on the functional form of these forces (Lakoba 2005, Molnar 1996, Parisi 2005, Yu 2005). We give some basic consideration to define the minimal complexity of the functional form of the repulsive forces together with some estimates of the values of parameters. From these considerations it becomes obvious that the repulsive forces have to depend not only on the relative position of persons, but also on the speeds and speed differences. The parameters of these forces will be situation dependant. They can in principle be derived from video observations of people moving, although the large scatter of data and the complexity involved makes for large uncertainties.

## 1. Introduction

The motion of pedestrians (e.g. for simulation of an evacuation situation) can be modelled as a multi-body system of self driven particles with repulsive interaction. Such modelling is important in the planning of large shops or offices, sports arenas and public transportation buildings. While many questions regarding safety (evacuation times) can be answered using the bottleneck capacity estimates or cellular automata models, these easy to use models are less suited for difficult situations and for level of service estimates. Here continuous space models have the capacity of becoming a universal tool for planning pedestrian facilities and are giving fine resolution estimates of individual movements. The details of such a model are still subject to discussion, and they have parameters which have to be adjusted correctly to give proper predictions of evacuation times, local densities, and forces on rails or obstacles. The social force model (Molnar and Helbing) in its various flavours is the oldest and most widely used continuous space model, and has shown to give good qualitative and quantitative predictions for some situations. We can show, however, that the functional form of repulsive force as defined there is not able to describe the full range of interactions correctly. We give an improved form of the repulsive force that can give good predictions for a much wider range of movements. For the single lane movement, we are able to deduce the numerical values of the force from the fundamental diagram. This carries over to multi-lane situations with high density, where there is no freedom of choice of directions and almost no passing. For lower densities and fairly free movement in a plane, we indicate how to estimate the steering, and get movements that are more realistic than those of the social force model. There is, however, still a substantial lack of data, such that the forces acting in this situation are not known with sufficient accuracy. Improving the data base will be the topic of further research in many places.

Our model is intended for use in situations where contact between pedestrians does not transmit substantial forces. Modelling a pushing crowd is a different story, and there is neither a model nor accurate data available at the time being.

## 2. Single Lane Movement

The movement in single lane is the simplest and therefore easiest to understand. On the other hand, many features of multi-lane movement and free movement in a plane can be described by single lane movement plus orientational behaviour, so single lane movement is basic to all pedestrian movement.

The normal movement of pedestrians involves physical forces between a person and the ground only, no forces between pedestrians. Therefore, all forces moving a person are willfully exercised by that person, and therefore depend on this persons perception of the situation. It is convenient to consider accelerating and decelerating influences seperately, as we will do here. With the mass of pedestrians set to unity, the assumption is

$$dv/dt = F_{de} + F_{acc} . \tag{1}$$

The accelerating force is generally [1] described as

$$F_{acc} = C_{acc} \cdot (v_{des} - v). \tag{2}$$

This expresses the fact that in a certain situation, a person has a desired speed $v_{des}$, and accelerates to this if not impeded. $C_{acc}$ and $v_{des}$ depend on the person and the situation, e.g. th espeed of walking up to a bus stop will increase as the departure time comes near. For modelling, a distibution of $C_{acc}$ and can be used. The constant is such that a person reaches about half the desired speed after one step, giving $F_{acc} \approx 1.6 \cdot (v_{des} - v)$, and $v_{des} \approx 1.36$ m/s for most situations in industrial countries. There is some scatter in the data, and while neither the functional form nor the values are fully accurate, the error is not larger than this scatter.

The decelerating force is more complicated, and neither functional form nor paramter values are agreed upon in literature. The first continuous space models of Molnar and Helbing used a fully symmetrical interaction force

$$F_{de} = F_{interact} = \Sigma_i \quad C_1 \cdot \exp(-|x_i - x|/C_2) \tag{3}$$

with $C_1 \approx 7$ , $C_1 \approx 0.3$ . This form is computationally inefficient, because persons at large distances give some contribution to it, however small. More important, it does not agree with common sense which tells us that we do not react on persons behind in normal walking. It has been replaced by more sophisticated forms without symmetry, e.g. getting the force as the gradient of a potential that is elongated in the direction of forward movement (Helbing).

The most general reasonable form of the interaction force is therefore

$$F_{de} = F_{interact} (x-x_1, v, v-v_1, \ldots, x-x_i, v-v_i) \tag{4}$$

where the force depends on the distance and the speed difference to a small number of persons in front and the persons own speed. We will see how much this can be simplified. Of course, the function $F_{interact}$ depends, like $C_{acc}$ and $v_{des}$ , on the individual and the situation.

### 2.1 Forces and the fundamental diagram

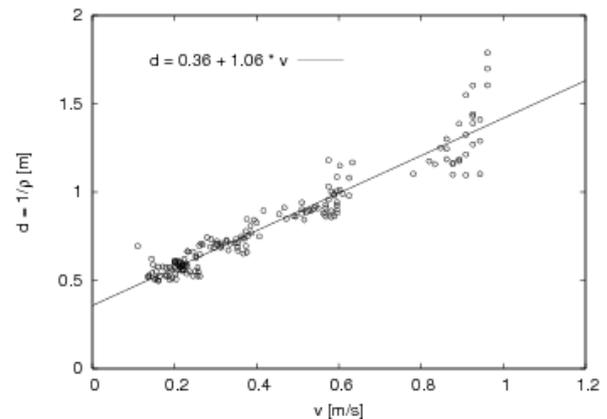

**Fig 1** Distance versus speed in single lane walking

The fundamental diagram gives a relation between density and flow (or speed) in a stationary situation. For single lane movement, instead of density per area, the density per length (or the mean distance $\Delta x$) is used.

The relation between the single lane fundamental diagram and the normal one for multi-lane unidirectional movement can be established via a (slightly speed dependent) lane width (Seyfried,2005). From and fitting, we get the relation

$$v = 0.94*\Delta x - 0.34$$

(5)     valid for $\Delta x = |x_{i+1}-x_i|$ from 0.36m to 1.5m for the situation measured, which is close to normal movement in offices or traffic facilities. For $\Delta x < 0.36m$ physical forces may be involved, and is nearly free movement and therefore of moderate interest. v will level off there.

At stationary flow, there is no acelleration, and

$$F_{interact} (x-x_1,v,0,\ldots,x-x_1,0) = -C_{acc} \bullet (v_{des}- v)$$

(6) It is further reasonable to assume that in the distances between persons further ahead are of no importance, thus with $x-x_i = \Delta x_i$ we get

$$F_{interact}(\Delta x,v(\Delta x),0,\Delta x_2,\ldots,\Delta x_i,0) = -C_{acc}\bullet(v_{des}-v(\Delta x))$$

(7) for any reasonable set of values of $\Delta x_2,\ldots,\Delta x_i$. Assuming now, like (3), that F does not explicitely depend on v, from (5) we get

$$F_{interact}(\Delta x) = -C_{acc} \bullet (v_{des} + 0.34 - 0.94\bullet\Delta x)$$

(8) For high density this is close to (2), but from $\Delta x = 0.5m$ on, the force is higher up to $\Delta x = 1.85m$. $F_{interact}$ becomes zero at $\Delta x \approx 1.81m$, indicating unimpeded movement there. For $\Delta x > 1.81m$ it will be set to 0.

## 2.2 Head On Colisions

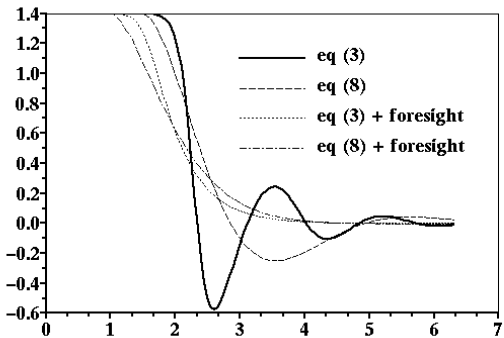

**Fig 2** Speed (m/s) versus time (s) for force (3), (8), and (3) and (8) with one second foresight.

Another situation accessible to analytic calculations is the head on collision of two persons. In this situtation, the force must be strong enough to stop both at a reasonable distance from each other. Further, the speed should come to zero smoothly, especially not show any oscillations. Neither (3) nor (8) fulfill this requirement, but stop too late and give oscillations in the final state. Further, it is easy to estimate that any interaction function depending on $\Delta x$ only will either be too strong to give a reasonable speed for walking in file at $\Delta x=1.5m$ or too weak to handle a head-on collision. Adding a dependance on v alone does not help, because due to (7), any such dependance will be compensated by a corresponding change of the dependance on $\Delta x$.

The most simple functional form possible will therefore depend on $\Delta x$ and on $\Delta v$. A simple and reasonable assumption is that a person does not react on the momentary situation, but has some foresight and therefore reacts on the extrapolation of the momentary situation. Maybe more realistic, but more complicated, would be an extrapolation out of the recent past to allow for reaction times. However, in the situations considered here, the difference will be too small to be detected, so we only try the simple extrapolation. Taking 1 s extrapolation time as a first guess we get

$$F_{de}(\Delta x, \Delta v) = -C_{acc} \bullet (v_{des} + 0.34 - 0.94\bullet(\Delta x-\Delta v\bullet s))$$

(9) for short distance, and 0 if (9) gives a positive force. A similar consideration can be made for the force (3). In both cases do we get the desired behavior for a head on collision, so there is no need to introduce further complications. The stronger dependance on v relative to (3) and (8) damps any oscillations.

## 2.3 Superposition of forces in 1D

Helbing, Molnar and others suppose that interaction forces are additive. We consider this for the movement of many persons with constant speed and identical distances in single lane. Assuming that eq. (5) holds and that $F_{interact}(\Delta x) = 0$ for $\Delta x > 1.87m$, assuming further that

$$F_{interact}(\Delta x,v(\Delta x),0,\Delta x_2,\ldots,\Delta x_i,0) =$$

(10)     $$f(\Delta x)+ f(2\Delta x)+ f(3\Delta x)+ \ldots$$

where f is the person-person interaction function. With $v_{des}$ = 1.36 (average value), for 0.935m<$\Delta$x<1.81m we have only one person-person interaction contributing to $F_{interact}$, for 0.603m<$\Delta$x<0.905m we have two such contributions, etc. This leads to

$$f(\Delta x) = -2.72+-0.94 \cdot \Delta x,$$
$$f(\Delta x) = -0.94 \cdot \Delta x , \qquad 0.603m < \Delta x < 0.905m \qquad (11)$$
$$f(\Delta x) = 2.72-6.02 \cdot \Delta x, \qquad 0.361m < \Delta x < 0.603m$$
$$f(\Delta x) = 5.44-13.5 \cdot \Delta x, \qquad 0.301m < \Delta x < 0.361m$$

The first line is obvious, if 0.905m < $\Delta$x, there is only one person-person interaction contributing, because all others are more than 1.81m away. We get to shorter distences by considering $f(\Delta x) = F_{interact}(\Delta x, ,...) - f(2\Delta x)$, where 0.905m < 2$\Delta$x < 1.81m, etc. It just happens considering three or four persons at less than 1.81m results in the same function. We see that close interactions are weak again, absolutely contrary to common sense,and for $\Delta$x<0.45m they even have the wrong direction. So in single lane movement we can assume that normally only the closest interaction matters, and only the strongest person-person interaction matters.

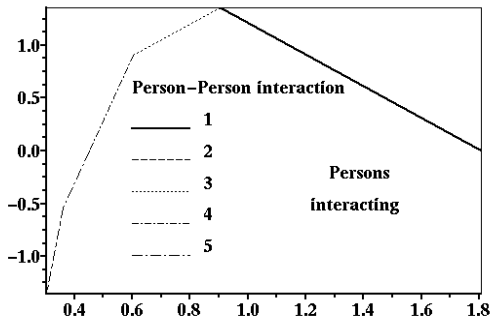

**Fig 3** Person-Person interactionforce

# 3. Unidirectional Multi-Lane Movement

In unidirectional movement in wider space at medium to high density, lane formation (Hoogedoorn 2005, Seyfried 2006) is the dominant structure. There are two modes of individual movement. The standard mode is just following the person ahead in the same lane, and the other – the less frequent the higher the density is – is changing the lane and moving into a neighbouring lane that allows a faster movement. The forces for the standard mode are governed by the same rules as the single lane movement. The changing of lanes is a 2D movement, and will be treated below. However, for the high densities that are of interest e.g. in evacuation studies, changing of lanes is too rare an event to have an effect on evacuation times. It may be, however, that the frequency of singular strong perturbations like people tripping or loosing baggage and stopping, is influenced by the frequency of line changes.

At low density, where passing is normally possible without problems, it can be assumed that all persons are moving at a speed close to $v_{des}$, and the influence of other persons affects the momentary direction to some extent, while the speed is reduced only so far as walking straight ahead  is  faster that taking a curve. The curve radii involved are quite large, so this effect is hardly detectable. It is not clear at which density lane formation sets in. Low density situations have not received much attention in the past, and only with the recent availability of automatic tracking from digital video images is this range accessible to experiments.

# 4. Movement in Two Dimensions

### 4.1 Seperation of steering and speed

The motion implied by eq. (3) is a sliding motion without preferred direction, that is the process of turnig the body is ignored. A more realistic model is one where every person is oriented, and the forward and turning movement are treated seperately, though not independantly.

The underlying concept is that – except in extreme density situations – a person walking will avoid obstacles primarily by moving around them, not by slowing down. Therefore any obstacle in the vicinity – other persons as well as objects - will be checked wether it is or will be in the pathway, and if it is, a turning force will be exerted. Only if there is no unobstructed reasonable path possible will there be a strong decellerating force. The reach of this 'turning force' is much larger than that of the decelleration force, such that normally the new direction is reached before the obstacle causes any slowing down. It is reasonable, though not essential, to assume that the process of turning – which is estimated to allow up to 90° per second – causes a little slowing down, but this effect is small unless the turning is really fast.

The figures show the simulated tracks of twenty people each entering a corridor through a door at random intervals with 1.5s mean and passing it, leaving through a door at the other end. Fig 4 shows a traditional modeling, where movement forward and sideward is on the same footing, although the force field is elongated in the direction of the present direction. In fig. 5, the persons are treated as oriented, and the turning movement sets in at three times the distance of the slowing down. Obviously, this results in much smoother and more realistic trajectories. There is one problem with this simulation: The turning force is assumed to depend smoothly on Δx, so that persons meeting exactly head-on are not turning much but slowing down. In the future, this will be replaced by a turning force where the direction will be chosen randomly at head on collisions, like people actually seem to do. The separation of orientation and speed in the movement leads to much smoother trajectories, while the times needed to pass the corridor are only marginally changed. Therefore the new approach does not so much improve the reliability of evacuation times as it improves level of service predictions.

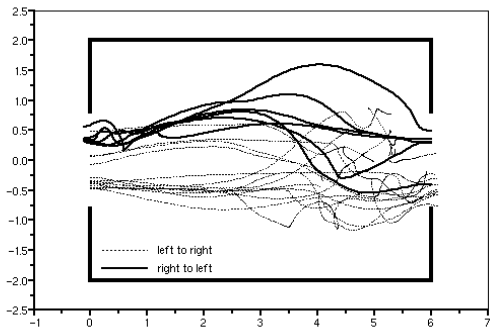

**Fig 4** Trajectories of passage through a corridor with force taken from Helbing, p.30

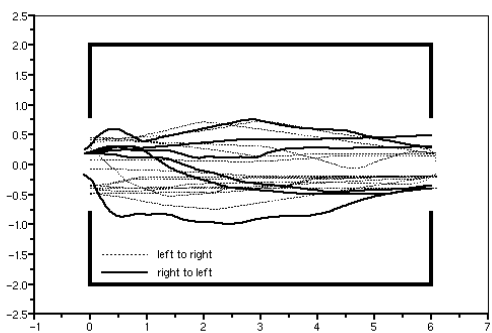

**Fig 5** Trajectories of passage through a corridor with turning and forward movement separated.

### 4.2 Superposition of forces in 2D

For the forces in longitudinal direction, there is no reason to assume any strong deviation from the 1D situation – the next encounter gives the longitudinal interaction force. For the lateral forces, which determine the direction of walking, there has to be some kind of superposition to avoid all collisions. The superposition should, however, not be simply addition. The steering will not even depend continuously on the situation, a pending head-on collision asks for strong steering either right or left, with little preference for on direction or the other. A proper rule might be: find the unobstructed path requiring the least amount of steering. The details of this are not yet settled.

## 5. Conclusions

The continuous space modeling of pedestrian movements has the potential of giving realistic individual trajectories. The motion is governed by forces that are exercised by all possible kinds of obstacles. The exact form of these forces is known with sufficient accuracy only for some important standard situations, but the knowledge is increasing. The original idea of symmetric additive forces is shown to be too simplistic. The concept of separating steering and forward motion greatly enhances the reliability of trajectories in medium and low density situations.